\documentclass[12pt,a4paper]{article}

\usepackage[a4paper,left=1.08in,right=1.08in,top=1.10in,bottom=1.10in,footskip=0.35in]{geometry}
\usepackage{setspace}
\AtBeginDocument{%
  \doublespacing
  \setlength{\textheight}{\dimexpr 27\baselineskip + \topskip\relax}%
}

\usepackage[T1]{fontenc}
\usepackage[utf8]{inputenc}
\usepackage{lmodern}
\usepackage{microtype}

\usepackage{amsmath,amssymb,amsfonts,amsthm,mathtools,bm}
\usepackage{graphicx}
\graphicspath{{Fig/}}
\usepackage{subcaption}
\usepackage{adjustbox}
\usepackage{threeparttable}
\usepackage{multirow}
\usepackage{booktabs}
\usepackage{xcolor}
\usepackage{algorithm}
\usepackage{algpseudocode}
\usepackage[toc,page]{appendix}
\usepackage[round,authoryear]{natbib}

\usepackage[colorlinks=true,allcolors=blue]{hyperref}
\usepackage[nameinlink,capitalise,noabbrev]{cleveref}

\theoremstyle{plain}
\newtheorem{theorem}{Theorem}
\newtheorem{assumption}{Assumption}

\newtheorem{corollary}{Corollary}
\newtheorem{assumption'}{Assumption}
\newtheorem{proposition}{Proposition}

\theoremstyle{remark}

\theoremstyle{definition}

\newcommand{\indep}{\perp\!\!\!\perp}
\algrenewcommand\alglinenumber[1]{\textbf{#1.}}

\newcommand{\reviewtitleblock}{%
  \begingroup
  \setstretch{1.1}
  \begin{center}
 {\Large\bfseries Assessing Estimate of CATE from Observational Data via an RCT Study\par}
  \vspace{0.35\baselineskip}
  {\normalsize Bosen Cui$^{1}$ and Yuhong Yang$^{1,2}$\par}
  \vspace{0.20\baselineskip}
  {\normalsize
  $^{1}$Yau Mathematical Sciences Center, Tsinghua University, Beijing 100084, China\\
  $^{2}$Beijing Institute of Mathematical Sciences and Applications, Beijing 101408, China\\
  Correspondence to: \href{mailto:yyangsc@mail.tsinghua.edu.cn}{yyangsc@mail.tsinghua.edu.cn}\par}
  \end{center}
  \vspace{0.35\baselineskip}
  \endgroup
}

\begin{document}

\reviewtitleblock

\begin{abstract}
Conditional average treatment effects (CATEs) are increasingly estimated from observational data and used to guide policy and individualized treatment decisions. Before such estimates can be trusted in practice, their predictive fitness needs to be assessed, yet observational data alone offer limited opportunities for doing so. We propose CATE Assessment via Fitness Evaluation (CAFE), a formal framework for directly assessing the goodness-of-fit of a CATE estimate learned from observational data, rather than the full underlying outcome model, using evidence from a randomized trial. CAFE partitions the trial covariate space according to estimated propensity scores (or the like) and compares observationally derived conditional treatment effects with group-level experimental averages. The framework accommodates a broad class of CATE learners, including parametric models and flexible machine learning methods such as causal forest and boosting. We establish theoretical guarantees under both the null and alternative hypotheses, and introduce a maximum-type extension to improve sensitivity to localized lack of fit. When both randomized trial and observational data are available, we further develop a two-stage procedure to detect the existence of unobserved confounders. Extensive numerical studies show the utility of the CAFE approach when assessing observational-derived CATE estimates.
\end{abstract}


\section{Introduction}
Precision medicine aims to tailor treatments to the characteristics of individual patients. Central to this goal is the concept of heterogeneous treatment effects (HTE), which recognizes that treatment efficacy may vary substantially across individuals. By accounting for such heterogeneity, clinicians can move beyond population-level average treatment effects (ATEs) toward individualized decisions that maximize patient benefit.
Motivated by this goal, substantial methodological progress has been made in estimating treatment effect heterogeneity \citep{cai2011analysis,imai2013estimating,athey2016recursive,taddy2016nonparametric,hahn2020bayesian}. A central estimand in this literature is the conditional average treatment effect (CATE), which measures the expected difference between potential outcomes under treatment and control, conditional on individual covariates. CATE estimation has become a foundational tool in both statistical methodology and applied research \citep{wager2018estimation,semenova2021debiased,yadlowsky2021estimation,kennedy2023towards}. 

Despite rapid progress in estimation methodology, a fundamental question remains insufficiently understood: \textit{How reliable is a CATE estimate learned from observational data?}  Overestimating heterogeneity may encourage overly aggressive personalization and expose individuals to ineffective or harmful treatments, whereas underestimating heterogeneity may reinforce a one-size-fits-all approach and obscure opportunities for targeted care \citep{van2019models}.
Existing approaches to uncertainty assessment for CATE estimation do not fully address this problem. Under unconfoundedness, several works study model selection or model averaging for CATE estimation by comparing candidate learners through  different validation criteria  \citep{rolling2014model,schuler2018comparison,alaa2019validating,rolling2019combining}.
These methods are useful for selecting or aggregating competing CATE estimators, but they do not provide a formal statistical procedure for assessing whether a given CATE model is correctly specified from a causal perspective.

This motivates the problem of goodness-of-fit assessment for CATE estimates trained on observational data. Classical goodness-of-fit tests are not designed for this task. Most existing procedures target standard regression or classification models \citep{fan2001goodness,shah2018goodness,jankova2020goodness,zhang2023classification,javanmard2024grasp}, where model adequacy is evaluated against the observed joint distribution of $(\mathbf{X},Y)$. In causal inference, however, the target involves a counterfactual object whose identification depends on additional assumptions, most notably the absence of unobserved confounding. Since these assumptions cannot be verified from observational data alone, and individual treatment effects are never directly observed, the existing goodness-of-fit tests in the literature cannot determine whether an observationally estimated CATE is causally valid.

At the same time, it is increasingly common to observe both a large observational study (OS), which supports flexible estimation but may suffer from hidden bias, and a smaller randomized controlled trial (RCT), which is internally valid but often limited in size and representativeness \citep{deaton2018understanding}. This dual-data structure raises a natural question: can RCT data be used to formally assess the goodness-of-fit of CATE estimates trained on observational data?
In this paper, we answer 
this question by
proposing CATE Assessment via Fitness Evaluation (CAFE), a formal hypothesis testing framework for the task. We consider a common but underexamined setting in which a flexible CATE model is learned from a large OS, potentially subject to unobserved confounding, and then evaluated against a smaller RCT that serves as an external benchmark. 

Our approach can go beyond a single global goodness-of-fit conclusion. A detected lack of fit may arise from at least two distinct sources: misspecification of the observational CATE model, or failure of the identifying assumptions required for causal interpretation in the OS. To isolate the possible reasons of lack of fit, we develop a two-stage testing procedure when both OS and RCT data are available. This yields an interpretable diagnostic tool that helps practitioners determine whether a rejection is attributable primarily to modeling deficiencies or to a more fundamental breakdown of causal identification. For completeness, we discuss related work and its connection to the proposed framework in the supplementary material.
\section{Assessing  CATE estimate from an observational study}\label{sec2}
\subsection{Potential outcome framework}
In this paper, we adopt the potential outcome framework \citep{neyman1923application, rubin1974estimating} 
and consider a setting with a binary treatment variable $A \in \{0,1\}$, observed covariates $\mathbf{X}\in \mathcal{X} \subseteq \mathcal{R}^p$, and an outcome of interest $Y \in \mathcal{R}$. Under the stable unit treatment value assumption (SUTVA), let $Y(a)$ represent the potential outcome when the subject receives treatment $a$, for $a \in \{0,1\}$. The observed outcome is given by: $Y =Y(A) = A Y(1)+ (1-A)Y(0)$. The outcome mean function is defined as $\mu_a (\mathbf{x}) =  E(Y(a) \mid \mathbf{X} = \mathbf{x})$.
Let $\tau(\mathbf{X}) := \mu_1(\mathbf{X}) - \mu_0(\mathbf{X}) = {E}[Y(1) - Y(0) \mid \mathbf{X}]$ denote the target conditional average treatment effect. In observational studies, identifying $\tau(\mathbf{X})$ typically requires the unconfoundedness assumption, $\{Y(1),Y(0)\} \indep A \mid \mathbf{X}.$
\subsection{Set-up of the observational study}
We consider an independent and identically distributed (i.i.d.) observational sample $\Delta_O=\{(\mathbf{X}_i,A_i,Y_i)\}_{i\in\mathcal{I}_m}$,  where
$\mathcal{I}_m$  indexes the $m$ units. Suppose we are provided with pre-trained estimates based on this dataset, including an estimate of 
$\tau(\mathbf{x})$, denoted as $\hat{\tau}_O(\mathbf{x})$ and a propensity score estimate $\hat{e}_O(\mathbf{x})$. These estimates may be obtained from either parametric models or flexible machine learning methods such as random forests or gradient boosting. Our primary goal is to assess the reliability of $\hat{\tau}_O$.


This setup reflects a common modern workflow in which fitted causal models or prediction functions are available, while the original individual-level data or model-training details may be inaccessible due to privacy, regulatory, or proprietary constraints. We therefore treat $\hat{\tau}_O(\mathbf{x})$ and $\hat{e}_O(\mathbf{x})$ as black-box inputs: they can be evaluated at new covariate values, but are not retrained, refitted, or audited in our procedure.
\subsection{Is
	the observational estimate $\hat{\tau}_O(\mathbf{x})$ valid for CATE?}
The validity of $\hat{\tau}_O(\mathbf{x})$ for estimating CATE depends on two key requirements. First, the observational study must satisfy the key causal identification condition of unconfoundedness, under which treatment assignment is conditionally independent of the potential outcomes given covariates. If this condition fails because of unobserved confounders, the CATE is generally not identified from the observed distribution of
$(\mathbf{X},A,Y)$.
Second, even under unconfoundedness, $\hat{\tau}_O(\mathbf{x})$ must be estimated appropriately.  For parametric methods, reliability is tightly linked to whether the model is correctly specified: if the underlying functional form is well matched to the true data-generating process, the estimator performs well; however, any form of misspecification can lead to systematic biases and unreliable estimates, regardless of sample size. Flexible nonparametric or machine learning methods reduce reliance on explicit functional-form assumptions, but their reliability still depends on sample size, tuning, regularization, and the stability of nuisance-function estimation. Thus, observational CATE estimates may be unreliable either because the causal identification assumptions fail or the estimation procedure itself performs poorly.


The above consideration underscores the importance of carefully evaluating the performance of $\hat{\tau}_O(\mathbf{x})$ in practice, rather than taking its validity for granted.
Such evaluation is difficult using observational data alone, because unconfoundedness is fundamentally unverifiable and individual treatment effects are never directly observed \citep{imbens2015causal}. External validation is therefore essential. Randomized controlled trials provide a natural benchmark: randomization balances observed and unobserved confounders in expectation and permits unbiased estimation of average causal effects. Independently collected RCT data can therefore be used to assess whether an observationally trained CATE estimate is compatible with randomized evidence. 
\subsection{Set-up of the randomized controlled trial}
Let $S \in \{0,1\}$ denote the indicator of study source, where $S=1$ corresponds to the RCT data and $S=0$ to the observational data. Define the respective propensity score as $e_{S}(\mathbf{X}) = P(A=1 \mid \mathbf{X},S)$. The covariate distributions in the RCT and OS are denoted by $p_R$ and $p_O$, respectively, and may differ. {
	Moreover, for each $a\in\{0,1\}$, we decompose the potential outcome as 
	$Y(a)=\mu_a(\mathbf{X})+\varepsilon_a$,
	where $E(\varepsilon_a\mid \mathbf{X})=0$ and $\mathrm{Var}(\varepsilon_a\mid \mathbf{X})=\sigma_a^2 >0$ almost surely. For simplicity, we assume homoskedastic errors. Extending the analysis to the heteroskedastic case is conceptually straightforward, although it requires additional technical derivations.} 
Let $\Delta_R=\{(\mathbf{X}_i,A_i,Y_i)\}_{i\in\mathcal{I}_n}$ denote the RCT sample of $n$ i.i.d. units. The following assumptions characterize the internal validity and relevance of the RCT:
\begin{assumption}	[RCT validity] \label{a1}
	(i) $\{Y(0), Y(1)\} \indep A \mid S=1, \mathbf{X}$, and (ii) for any $\mathbf{x}\in\mathcal{X}$, we have $0<e_1(\mathbf{x})<1$. 
\end{assumption}
\begin{assumption}[Transportability of the CATE]\label{a2}
	There exists a function $\Theta(\mathbf{Z})$ such that 
	\[
	E[Y(1) - Y(0) \mid \mathbf{Z}, S = 1] = E[Y(1) - Y(0) \mid \mathbf{Z}, S = 0] = \Theta(\mathbf{Z}).
	\]
	Here, $Z$ denotes the set of treatment effect modifiers, which may include some components of $X$, and possibly some unobserved covariates. 
\end{assumption}
Assumption \ref{a1}(i) reflects the randomized nature of the trial and ensures internal validity; Assumption \ref{a1}(ii) guarantees that all subjects have positive probabilities of receiving each treatment. In most cases, $e_1(\mathbf{X})$ is known in RCT design. Assumption \ref{a2} is a form of CATE transportability—it allows the treatment effects identified in the RCT population to be extended to the OS population, despite possible differences in baseline covariate distributions. 
This assumption is weaker than the CATE transportability conditions imposed in, for example, \cite{lee2023improving,cole2010generalizing,colnet2024re}. In particular, it permits unmeasured confounding in the OS, provided that the latent effect-modifying structure governing treatment heterogeneity is shared across the two data sources.


In the remainder of this paper unless otherwise stated, we require the support of the RCT covariates to coincide with that of the OS covariates, so that $\hat{\tau}_{O}(\mathbf{x})$ is well defined for RCT units. This common-support condition is standard in the data-fusion literature \citep{kallus2018removing,cheng2021adaptive,wu2022integrative}.

\section{CATE assessment via fitness evaluation (CAFE)} \label{sec3}
To evaluate the fitness of a given CATE estimate $\hat{\tau}_O(\mathbf{x})$,  we propose CATE Assessment via Fitness Evaluation (CAFE), a two-step procedure consisting of a partitioning step and a testing step. Direct validation of individual-level treatment effect estimates is generally difficult, whereas subgroup-level treatment effects can be estimated more stably from randomized data. CAFE exploits this fact by partitioning the RCT covariate space and testing whether the group-aggregated predictions from $\hat{\tau}_O$ are compatible with the corresponding experimental evidence.
\subsection{Partitioning step}
We partition the RCT covariate space $\mathcal{X}$ into $K_n$ disjoint groups, denoted by  $\mathcal{P} = \{P_{R,1},\ldots,P_{R,K_n}\}$. These groups act as strata for comparing the estimated effects with those observed experimentally.
The partition is required to satisfy the following conditions: First, conditional on covariates, the partitioning rule should be independent of the potential outcomes $\{ Y(0),Y(1)\}$. Second, the number of groups $K_n$ may increase with the sample size but at a controlled rate, ensuring that each group contains a sufficient number of observations. The above requirements ensure that the partitioning procedure does not introduce post-treatment bias and preserves the validity of causal comparisons within each group. 

In our general framework, we allow any partition $\mathcal{P}$ that adheres to these principles.
In the following sections, unless otherwise specified, we use propensity score stratification based on $\hat{e}_O(\mathbf{x})$ as a running example. In the supplementary material, we provide detailed descriptions of alternative partitioning strategies and report additional simulation results.
\subsection{Test statistic}
Given a partition $\mathcal{P}$, we compare the observational estimate $\hat{\tau}_O(\mathbf{x})$
with RCT-based group-level treatment effect estimates. Let $\tau_{R,k} = \frac{1}{n_k} \sum_{i: \mathbf{X}_i \in P_{R,k}} \tau(\mathbf{X}_i)$ denote the conditional group-specific average treatment effect for group \(P_{R,k}\),
given the realized RCT covariates,  and let $\hat{\tau}_{R,k}$ be its difference-in-means estimator. Let $n_k=\sum_{i=1}^n \mathbb{I}\{\mathbf{X}_i \in P_{R,k}\}$ be the group size. We define the CAFE statistic as
\begin{equation*}
	T = \sum_{k=1}^{K_n} \left(\frac{  \sum_{i: \mathbf{X}_i \in P_{R,k}}\left[\hat{\tau}_{R,k}-\hat{\tau}_O(\mathbf{X}_i)\right]}{n_k\sqrt{\hat{\sigma}_k^2}}\right)^2,
\end{equation*}
where $\hat{\sigma}_k^2$ is a consistent estimator of 
$\sigma_k^2=\text{Var}(\hat{\tau}_{R,k})$. { In particular, letting $n_{1,k}$ and $n_{0,k}$ denote the numbers of treated and control units in $P_{R,k}$, respectively,
	we estimate $\sigma_k^2$ by the sample-variance plug-in form
	\begin{equation*}
		\hat{\sigma}_k^2=\frac{s_{1,k}^2}{n_{1,k}}+\frac{s_{0,k}^2}{n_{0,k}},
	\end{equation*}
	where $s_{a,k}^2$ is the sample variance of outcomes among units with treatment $A=a$ within $P_{R,k}$.}

Under the null hypothesis that $\hat{\tau}_O$ is a valid estimate of CATE, $T$
is calibrated against a chi-squared distribution with $K_n$ degrees of freedom. The corresponding $p$-value is \begin{equation*}
	P(\chi^2_{K_n} > T\mid T,K_n),
\end{equation*}
and the null is rejected when the $p$-value is smaller than a pre-specified significance level $\alpha$. 



\subsection{Assessing parametric CATE estimators}

Parametric CATE models assume a specific functional form for how the treatment effect depends on covariates. Fitness testing in this context reduces to checking whether the  CATE function belongs to the proposed parametric family: $$\tau(\cdotp) = f(\cdotp;\beta),$$ where $f$ is known and $\beta$ (unknown) is in a  finite dimensional parameter space $\mathcal{B} \subseteq \mathcal{R}^p$.
The null and alternative hypotheses for goodness-of-fit testing of a parametric CATE model are defined as \begin{equation} \label{H0_para}
	H_0: \tau(\cdotp) \in \{f(\cdotp,\beta)|\beta \in\mathcal{B}\}, 
\end{equation} 
and 
\begin{equation}\label{H1_para}
	H_1: \tau(\cdotp) \notin \{f(\cdotp,\beta)|\beta \in\mathcal{B}\}.
\end{equation}
{Under this formulation, the null hypothesis asserts that the proposed parametric specification is compatible with the true treatment effect. When the test detects incompatibility, it may reflect either misspecification of the parametric form for $\tau(\cdot)$, or failure of the identifying assumptions, such as the existence of unobserved confounding.}
Note that even when $\tau(\cdot)$ is modeled parametrically, nuisance components used to construct $\hat{\tau}_O(\cdot)$ may still be estimated using semiparametric or nonparametric techniques.
\subsection{Assessing general CATE estimation procedures}
Modern CATE estimation increasingly relies on nonparametric and machine-learning approaches such as random forests, boosting and meta-learners that do not rely on fixed finite-dimensional models \citep{wager2018estimation,kunzel2019metalearners,chen2016xgboost,chernozhukov2018double,zhao2021causal}. 
A central question for such general CATE procedure is how to assess whether the proposed method provides a sufficiently good estimate, as reflected in its convergence to the CATE $\tau(\mathbf{x})$?

Let $r_m$ denote a possible  convergence rate of the observational estimator $\hat{\tau}_O(\mathbf{x})$. The null hypothesis is \begin{equation}\label{H0_gen}
	H_0: \space \sup_{\mathbf{x} \in\mathcal{X}} |\hat{\tau}_{O}(\mathbf{x}) -\tau(\mathbf{x})| = O_p(r_m).
\end{equation}
The alternative hypothesis is that there exists  $M_n \subset \mathcal{X}$ with $P(\mathbf{x} \in M_n)$ bounded away from 0 such that  \begin{equation}\label{H1_gen}
	H_1: \inf_{x\in M_n} |\hat{\tau}_{O}(\mathbf{x}) -\tau(\mathbf{x})| /r_m \stackrel{p}{\to} \infty.
\end{equation}
Thus, under the null, the observational estimator achieves the prescribed uniform convergence rate, whereas under the alternative, it fails by converging at a slower rate on a non-negligible region of the covariate space. This formulation accommodates a broad class of modern CATE estimators. For instance, sieve-based debiased or double machine learning estimators often yield uniform stochastic errors of order $\sqrt{\frac{\log m }{m}}$, up to approximation and complexity terms \citep{semenova2021debiased}. In reduced-dimensional local-smoothing frameworks, rates of the form  $(m/\log m)^{-s/(2s+d)}$, may be obtained, where $s$ indexes smoothness of $\tau(\mathbf{x})$ and $d$ the dimension of the effect-modifying covariates \citep{fan2022estimation}.
\section{Asymptotic results for CAFE} \label{sec4}
\subsection{Type I error of CAFE}
We first establish the asymptotic validity of CAFE under the null hypotheses. For any positive sequences $a_n$  and $b_n$, $a_n =\omega(b_n)$ if $\frac{a_n}{b_n} \to\infty$, $a_n = \Omega (b_n)$ if there exists a constant $C>0$, such that $\frac{a_n}{b_n} \ge C$, as $n\to\infty$. The following assumptions impose regularity conditions on the RCT partitions and the outcome distribution.
\begin{assumption} 	[Sufficient observations in each group]\label{a3}
	There exists a positive sequence $\{c_n\}$ such that $\min_{k=1,\ldots,K_n} \sum_{i\in\mathcal{I}_n}\mathbb{I}(\mathbf{X}_i \in P_{R,k})\ge c_n$ a.s., $c_n = \omega (n^{5/7})$ as $n\to\infty$. There exists a constant $0<\eta<\frac{1}{2}$ such that $\eta \le P(A_i =a | X_i\in P_{R,k})\le 1- \eta$ a.s.  
\end{assumption}
{\begin{assumption}[Bounded within-group heterogeneity and variance] \label{a4}
		There exists a positive constant $\zeta_1$, such that
		\begin{equation*}
			\sup_{1\le k \le K_n} \sup _{\mathbf{x} \in P_{R,k}}|\tau_{R,k} - \tau(\mathbf{x})| \le \zeta_1.
		\end{equation*}
		And there exists constants $0<\zeta_2 \le \zeta_3<\infty$ such that \begin{equation*}
			\zeta_2\le\inf_{1\le k\le K_n} n_k \sigma_k^2 \le \sup_{1\le k \le K_n} n_k \sigma_{k}^2 \le \zeta_3, \quad a.s.
		\end{equation*}
	\end{assumption}
	\begin{assumption} 	[Finite eighth moment for potential outcome]\label{a5} The potential outcomes have bounded eighth moments; that is
		\begin{equation*}
			E \left[ Y\left(1\right)^8\right]+ E \left[ Y\left(0\right)^8\right] \le C,
		\end{equation*}
		for some constant $C>0$.
\end{assumption}}

Assumption \ref{a3} requires a minimum number of observations in the partitioned RCT group. This is guaranteed by requiring the group sizes to grow at a rate $c_n = \omega(n^{5/7})$, and can be achieved in practice by selecting appropriate quantile-based partitions of the estimated propensity score derived from the observational data. Additionally, the assumption enforces a uniform positivity condition on treatment assignment probabilities, ensuring identifiability within each group.
{Assumption \ref{a4}  rules out excessive within-group heterogeneity and degenerate variance, while Assumption \ref{a5}  imposes a mild moment condition used for asymptotic approximation.

	The following theorem shows that under the null hypothesis $H_0$
	the CAFE test statistic follows a chi-squared distribution asymptotically. 
	\begin{theorem} [Type I error for a general learner]\label{th1}
		Assume that Assumptions \ref{a1}-\ref{a5} hold, under $H_0$ (\ref{H0_gen}), if $n \to\infty$ and $n = o(r_m^{-14/9})$ as $n\to\infty$, we have $ P(\chi^2_{K_n} > T\mid T,K_n)\stackrel{d}{\to}U$, where $U\sim \text{Unif} (0,1)$.
	\end{theorem}
	For parametric CATE estimates, we modify the convergence condition as follows:
	\begin{assumption}	[Parametric rate of convergence under $H_0$] \label{a7}
		Under $H_0$, as $m\to \infty$,  \begin{equation*}
			\sup_{\mathbf{x} \in\mathcal{X}} |\hat{\tau}_O(\mathbf{x}) - \tau(\mathbf{x})| = O_p(1/\sqrt{m}). 
		\end{equation*} 
	\end{assumption}
	\begin{theorem} [Type I error for a parametric model] \label{th2} 
		Assume that Assumptions \ref{a1}-\ref{a7}  hold,  under $H_0$ (\ref{H0_para}), if $n\to\infty$ and $n = o(m^{7/9})$ as $n \to \infty$,  we have $ P(\chi^2_{K_n} > T\mid T,K_n)\stackrel{d}{\to}U$.
	\end{theorem}
	Theorems \ref{th1} and \ref{th2} show that the CAFE $p$-value is asymptotically valid under $H_0$. The restriction on $n$ in the above theorems can be easily achieved in our setting. When a nonparametric method is used, the slower convergence rate requires the observational sample size $m$ to be significantly larger to ensure the validity of the test.
	
	\subsection{Power of CAFE}
	We next study the power of CAFE under $H_1$. Below, we introduce some necessary assumptions for the theorem.
	
	\begin{assumption}[General identifiable difference under $H_1$]\label{a8}
		Under $H_1$ (\ref{H1_gen}), with probability going to one, there exists $M_n \subseteq \mathcal{X}$, such that either \begin{equation*}
			\text{ess} \inf_{\mathbf{x} \in M_n} (\hat{\tau}_O(\mathbf{x})-\tau(\mathbf{x})) \ge 0,
		\end{equation*}
		or \begin{equation*}
			\text{ess} \sup_{\mathbf{x}\in M_n} (\hat{\tau}_O(\mathbf{x})-\tau(\mathbf{x}))\le 0,
		\end{equation*}
		and \begin{equation*}
			\inf_{\mathbf{x} \in M_n} \frac{
				\left| \hat{\tau}_O(\mathbf{x})-\tau(\mathbf{x}) \right|}{r_m^{(a)}} \ge \zeta_4 \quad a.s.
		\end{equation*}
		where $r_m^{(a)} \to 0$, $r_m^{(a)} = \omega (r_m)$ and constant $\zeta_4>0$. There is at least one group indexed by $k^*$ with
		\begin{equation*}
			P\left[\frac{\underline{n}_{k^*}}{n_{k^*}} > \zeta_5 \right] \to 1,
		\end{equation*}
		where $\underline{n}_{k^*} = \sum_{i=1}^n \mathbb{I}\{\mathbf{X}_i \in P_{R,k^*}\cap M_n \}$ and constant $0<\zeta_5<1$. Furthermore, 
		\begin{equation*}
			\sup_{\mathbf{x}\in\{ P_{R,k^*} \backslash M_n\}}\left|\tau(\mathbf{x}) - \hat{\tau}_O (\mathbf{x})\right|< \frac{\zeta_4 \zeta_5}{1-\zeta_5}r_m^{(a)}, a.s.
		\end{equation*} 
	\end{assumption} 
	Assumption \ref{a8} imposes an identifiability condition under the alternative hypothesis $H_1$. It ensures that, with high probability, there exists a non-negligible region $M_n\subseteq \mathcal{X}$ where the discrepancy between the observationally estimated CATE  and the true CATE is detectable. The condition can typically be met if $H_0$ is violated significantly and the number of groups is not too large. 
	
	\begin{theorem} [Consistency of general CATE estimation method]\label{th3}  Under $H_1$ (\ref{H1_gen}), assume Assumptions \ref{a1}-\ref{a5} and \ref{a8} hold, $n\to\infty$ and $n= \Omega((r_m^{(a)})^{-14/3})$. Then we have \begin{equation*}
			P(\chi^2_{K_n}\le T \mid T,K_n) \stackrel{p}{\to} 1.
		\end{equation*}
	\end{theorem}
	Combining Theorem \ref{th1} and \ref{th3}  yields the following corollary.
	\begin{corollary}
		Assume that $r_m^{(a)} = \Omega(r_m^{\lambda})$ as $n\to \infty$ with $0< \lambda < 1/3$, and Assumptions \ref{a1}-\ref{a5} and \ref{a8} hold, respectively. If $n$ satisfies $n=o (r_m^{-14/9})$ and $n = \Omega (r_m^{-14\lambda/3})$, 
		then CAFE has asymptotically valid Type I error under $H_0$ and power tending to one under $H_1$.
	\end{corollary}
	For parametric CATE estimate, we use the following alternative assumption.
	\begin{assumption}[Convergence and identifiable difference for parametric model under $H_1$]\label{a9} 
		Under $H_1$ (\ref{H1_para}), there exists a function $\tau_c : \mathcal{X} \to \mathcal{R}$  not belonging to the parametric family $\{f(\cdot,\beta) \mid \beta\in \mathcal{B}\}$ such that \begin{equation*}
			\sup_{\mathbf{x} \in \mathcal{X}} \left\vert \hat{\tau}_O(\mathbf{x})-\tau_c(\mathbf{x})\right\vert \stackrel{p}{\to} 0
		\end{equation*}
		when $m \to \infty$.
		With probability going to one, there exists $M_n \subseteq \mathcal{X}$, such that \begin{equation*}
			\text{ess} \inf_{\mathbf{x} \in M_n} ({\tau}(\mathbf{x})-\tau_c(\mathbf{x})) \ge c,
		\end{equation*}
		or \begin{equation*}
			\text{ess} \sup_{\mathbf{x}\in M_n} ({\tau}(\mathbf{x})-\tau_c(\mathbf{x}))\le -c,
		\end{equation*}
		for a positive constant $c$ and there is at least one group indexed by $k^*$ such that 
		\begin{equation*}
			P\left[\frac{\underline{n}_{k^*}}{n_{k^*}} > \zeta_6 \right] \to 1,
		\end{equation*}
		where $\zeta_6 <1$ is a positive constant. Furthermore,
		\begin{equation*}
			\sup_{\mathbf{x}\in\{ P_{R,k^*} \backslash M_n\}} \left| \tau(\mathbf{x})-\tau_c(\mathbf{x})\right|<  \frac{c\zeta_6}{1-\zeta_6}\quad a.s.
		\end{equation*}
	\end{assumption}
	\begin{theorem} 	[Consistency of parametric model]\label{th4}
		Under $H_1$ (\ref{H1_para}), assume Assumptions \ref{a1}-\ref{a5} and \ref{a9} hold, $n,m\to\infty$. Then we have \begin{equation*}
			P(\chi^2_{K_n}\le T \mid T,K_n) \stackrel{p}{\to} 1.
		\end{equation*}
	\end{theorem}
	Theorem \ref{th4} establishes the consistency of CAFE for parametric CATE estimation methods under the alternative hypothesis.
	
	\section{CATE assessment via fitness evaluation-maximum (CAFE-M)} \label{sec5}
	The CAFE statistic aggregates squared discrepancies across all partition groups and is therefore well suited to alternatives in which the lack of fit is spread over many groups. However, when the violation is localized to only a few groups, the signal may be diluted by many near-null components. To improve sensitivity to such localized departures, we introduce a maximum-type variant, CAFE-M, which focuses on the largest group-level standardized discrepancy.
	
	The CAFE--M test statistic is defined as  
	\begin{equation*}
		M = \max_{1\le k\le K_n} \left|\frac{  \sum_{i: \mathbf{X}_i \in P_{R,k}}\left[\hat{\tau}_{R,k}-\hat{\tau}_O(\mathbf{X}_i)\right]}{n_k\sqrt{\hat{\sigma}_k^2}}\right|.
	\end{equation*}
	Denote by $G$ a standard Gumbel random variable. The following results establish the Type-I error control and power consistency of CAFE-M.
	
	\begin{theorem}[Type I error of CAFE-M for general method]\label{th5} 
		Assume Assumptions \ref{a1}–\ref{a5} hold.  Under the null hypothesis $H_0$ (\ref{H0_gen}),
		if $n\to\infty$ with $n =o(\frac{1}{r_m^{2}|\log r_m|})$, then
		\[
		1-P\!\left\{ G \le \frac{M-a_{K_n}}{b_{K_n}} \;\mid\; M,K_n \right\}
		\;\xrightarrow{d}\; U,
		\]
		where
		$a_{K_n}= \Phi^{-1}(1-\frac{1}{2K_n})$, $
		b_{K_n}= \frac{1}{a_{K_n}},
		$
		and $\Phi(\cdot)$ is the standard normal cumulative distribution function.
	\end{theorem}
	By Theorem~\ref{th5}, the normalised statistic
	$(M-a_{K_n})/b_{K_n}$ is asymptotically standard Gumbel, and hence the resulting
	$p$–value is
	\[
	\hat{p}\;=\;1-\exp\!\bigl\{-e^{-(M-a_{K_n})/b_{K_n}}\bigr\}.
	\]
	Compared with the CAFE statistic, the null validity of CAFE-M requires only $n=o(\frac{1}{r_m^{2}|\log r_m|})$, which is substantially weaker than $n=o(r_m^{-14/9})$. Since it only requires uniform control of the largest deviation term at the extreme-value scale, CAFE-M is less demanding on the convergence rate of the observational estimate $\hat{\tau}_O$, making it more favorable when $r_m$ is slow, such as in high-dimensional or nonparametric learning problems.
	
	
	\begin{theorem}[Power of CAFE–M test]\label{th6} 
		Under the alternative hypothesis $H_1$ (\ref{H1_gen}), suppose Assumptions
		\ref{a1}–\ref{a5} and \ref{a8} hold and $n\to\infty$ with
		$n=\Omega\!\bigl((r_m^{(a)})^{-14/5} |\log r_m^{(a)}|^{7/5} \bigr)$. Then
		\[
		P\!\left\{ G \le \frac{M-a_{K_n}}{b_{K_n}} \;\mid\; M,K_n \right\}
		\;\xrightarrow{p}\; 1 .
		\]
	\end{theorem}
	When the misspecification is most serious in a local region and the grouping captures the defect in at least one group, CAFE--M can achieve consistency under weaker data requirements than CAFE.
	
	The parametric version follows by replacing the general convergence rate $r_m$ with the corresponding parametric rate and adapting the arguments from the previous section.

	
	\section{Attributing lack of fit when both RCT and OS data are available} \label{new sec}
	Rejection of the null hypothesis $H_0$ (\ref{H0_para}) by CAFE or CAFE-M via the RCT sample  provides strong evidence that the OS-based CATE estimate is unreliable. Such a rejection can arise from two different reasons:
	the presence of unobserved confounding in the observational study and/or inadequate performance of the learner in estimating the regression functions. When only  the  estimated CATE  $\hat{\tau}_O(\mathbf{x})$ and the propensity score are available, but not the original observational data itself, these two sources cannot be separated.
	
	When both the randomized and observational samples are available, however, additional diagnostics are possible. Define the Conditional Average Treatment Difference (CATD) as
	\begin{equation}\label{psi}
		\psi(\mathbf{x}) = E\left[Y \mid A=1, \mathbf{X}= \mathbf{x}\right] -  E\left[Y \mid A=0, \mathbf{X}= \mathbf{x}\right] = \tau(\mathbf{x}) + b(\mathbf{x}),
	\end{equation}
	where $b(\mathbf{x})$ is the bias function induced by unobserved confounding. 
	Under the  unconfoundedness assumption, $b(\mathbf{x}) = 0$, and the CATD coincides with the CATE $\tau(\mathbf{x})$. Without unconfoundedness, the observational distribution identifies $\psi(\mathbf{x})$, but $b(\mathbf{x})$ is not estimable without additional information.


	Building on the decomposition (\ref{psi}), we propose a two-stage testing procedure in which the first stage assesses whether the observational CATE is incompatible with RCT evidence, while the second stage examines whether this incompatibility is driven by unobserved confounding  and/or misspecification of the learner.
	For clarity, we describe the second stage for parametric CATD models. Consider \begin{equation*}
		H_{0}^{OS} : \psi(\cdot) \in \{f(\cdot,\beta) \mid \beta \in\mathcal{B}\},
	\end{equation*}
	and \begin{equation*}
		H_1^{OS} : \psi(\cdot) \notin \{f(\cdot,\beta) \mid \beta \in\mathcal{B}\},
	\end{equation*}
	where $\{f(\cdot,\beta) \mid \beta \in\mathcal{B}\}$ is the set of functions of CATD based on the parametric models under consideration.
	
	Suppose we have an additional observational test set of size $\tilde{n}$ and $\tilde{n} \asymp n$, which may either be obtained by sample splitting from the original observational dataset or collected independently from the same underlying data-generating process. Using CAFE as an illustrative example,
	partition the observational test set into $K_n$ disjoint groups $\mathcal{P}_O = \{P_{O,1},\ldots,P_{O,K_n}\}$ based on the same rule in stage 1 and compute \begin{equation*}
		T_O = \sum_{k=1}^{K_n} \left(\frac{  \sum_{i: \mathbf{X}_i \in P_{O,k}}\left[\hat{\tau}_{O,k}-\hat{\tau}_O(\mathbf{X}_i)\right]}{\tilde{n}_k\sqrt{\tilde{\sigma}_k^2}}\right)^2,
	\end{equation*}
	where $\hat{\tau}_{O,k}$ is the within-group difference-in-means estimator of the CATD, $\tilde{\sigma}^2_{k}$ is its consistent variance estimator, $\tilde{n}_k= \sum_{i=1}^{\tilde{n}} \mathbb{I}\{\mathbf{X}_i\in P_{O,k}\}$. The null hypothesis $H_0^{OS}$ is rejected if the corresponding
	$p$-value $p_o :=P(\chi^2_{K_n} > T_O \mid T_O, K_n)<\alpha$.

	The two-stage testing process lead to following possible decisions.\\
	\noindent\textit{$D_1$:}
	The randomized data provide no evidence against the compatibility of the observationally derived CATE estimator with the RCT benchmark; that is, the Stage 1 test fails to reject $H_0$.\\
	\noindent\textit{$D_2$:}
	The rejection of $H_0$ is attributed to unobserved confounding in the observational study; that is, $H_0$ is rejected but $H_0^{OS}$ is not rejected.\\
	\noindent\textit{$D_3$:}
	The rejection of $H_0$ is attributed to inadequate modeling of the CATD, while acknowledging that the presence of unobserved confounding cannot be ruled out, i.e., the null hypotheses in both stages are rejected.
	\begin{proposition}\label{pro1}
		Assume Assumptions \ref{a1}-\ref{a7} and \ref{a9} hold. If the sample sizes satisfy $n,m \to \infty$ with $n = o(m^{7/9})$, then, \begin{equation*}
			P(D_1 \mid H_0) \ge 1-\alpha+o(1), \quad\quad P(D_1 \mid H_1) {\to} 0. 
		\end{equation*}
		
		Furthermore, under similar assumptions of $T$ for the statistic $T_O$ (see the supplementary materials for details), we have \begin{equation*}
			P(D_2\mid H_1, H_0^{OS}) \ge 1-\alpha +o(1), \quad \quad P(D_2 \mid H_1, H_1^{OS})   \to 0,
		\end{equation*}
		\begin{equation*}
			P(D_3 \mid H_1, H_0^{OS}) \le \alpha+o(1), \quad\quad P(D_3 \mid H_1, H_1^{OS}) \to 1,
		\end{equation*}
		and \begin{equation*}
			P(D_2 \mid H_0)\le \alpha+o(1), \quad\quad P(D_3 \mid H_0) \le \alpha^2+o(1).
		\end{equation*}
	\end{proposition} 
	Proposition~\ref{pro1} establishes the asymptotic validity of the decisions by the proposed two-stage testing procedure. It consistently distinguishes compatibility, confounding-driven discrepancy, and model misspecification with small error probabilities.
	\section{Simulation} \label{sec6}
	We conduct simulation studies to evaluate the finite-sample performance of CAFE and CAFE-M. The proposed tests are applied to CATE estimates obtained from several commonly used learners, including the S-learner, T-learner, and R-learner \citep{kunzel2019metalearners,nie2021quasi}. Unless otherwise stated, the RCT and observational samples share the same potential outcome models and covariate distributions, but differ in treatment assignment. In the RCT, treatment is completely randomized with probability \(1/2\).
	
	\subsection{Parametric Settings}\label{parametric}
	We first consider three low-dimensional parametric designs. In each setting, we compare a correctly specified model with a misspecified model, and use the former to assess type I error and the latter to assess power.\\
	\textit{Parametric Setting 1 .} 
	We generate \begin{equation*}
		Y = \mathbf{X}^T \bm{\beta}_0 + A \tau(\mathbf{X}) + \epsilon,
	\end{equation*}
	where $\mathbf{X} = (X_1,\ldots, X_5)^T$, the covariates are independently generated from $U(0,5)$, $\bm{\beta}_0 = (1, -1, 0.5, 0, 1)^T$, $\tau(\mathbf{x}) = \mathbf{x}^T \bm{\beta}_1$, $ \bm{\beta}_1 = (0.5, 0.5, 0, -0.5, 1)^T$ and $\epsilon \sim \mathcal{N}(0,4)$. In the observational study, treatment is assigned according to
	\[
	e_0(\mathbf{x})=
	\frac{1}{1 + \exp(-\mathbf{x}^T \bm{\beta})},
	\quad
	\bm{\beta} = (0.5, -0.3, 0.2, 0.1, -0.1)^T.
	\]
	We estimate $e_0(\mathbf{x})$ via logistic regression and consider (i) a correctly specified CATE model and (ii) a misspecified model that omits $X_5$ from both the CATE and propensity score models.\\
	\textit{Parametric Setting 2.} We next consider interaction between the covariates \begin{equation*}
		Y =X_2+ A(3X_1- 2 X_2 + 0.5 X_1X_2)+\epsilon,
	\end{equation*}
	where $X_j\sim U(0,3)$, $\epsilon \sim \mathcal{N}(0,1)$. Propensity score in the observational studies is $\frac{1}{1+\exp\{-0.5x_1+0.3x_2\}}$.
	
	We again use logistic regression to estimate $e_0(\mathbf{x})$, testing both the correctly specified model and a misspecified model that omits the interaction term $x_1x_2$.\\
	\textit{Parametric Setting 3.} Consider nonlinear treatment effect \begin{equation*}
		Y = \mathbf{X}^T \bm{\beta}_0 + A\tau(\mathbf{X}) +\epsilon
	\end{equation*}
	where $\mathbf{X} = (X_1,\ldots, X_5)^T$, the covariates are independently generated from $\mathcal{N}(0,2)$ and $\epsilon \sim \mathcal{N}(0,1)$. Here,  $\bm{\beta}_0 = (1, -1, 0.5, 0.6, 1)^T$, $\tau(\mathbf{x}) = 0.3x_1+ 0.2 x_2-0.4x_3+0.2x_4-0.5x_5+x_1^2$. The observational study uses the propensity score $e_0(\mathbf{x})=
	\frac{1}{1 + \exp(-\mathbf{x}^T \bm{\beta})},$
	where $\bm{\beta} = (0.5, -0.3, 0.2, 0.1, -0.1)^T.$
	
	We estimate $e_0(\mathbf{x})$ using logistic regression and again contrast a correctly specified model with a misspecified model that omits the quadratic term $x_1^2$.
	
		
		Across the three settings, the observational sample sizes are \(m\in\{200,400,800\}\), with corresponding RCT sample sizes \(n\in\{40,80,120\}\). Each experiment is repeated 100 times. Propensity scores are estimated from the observational data, and the number of groups is set to \(K_n=\lfloor n^{2/7}\rfloor\). Additional simulation settings with dependent covariates and covariate shift are reported in the supplementary material. 
		\begin{figure}[h]
			\centering
			\includegraphics[width=\textwidth]{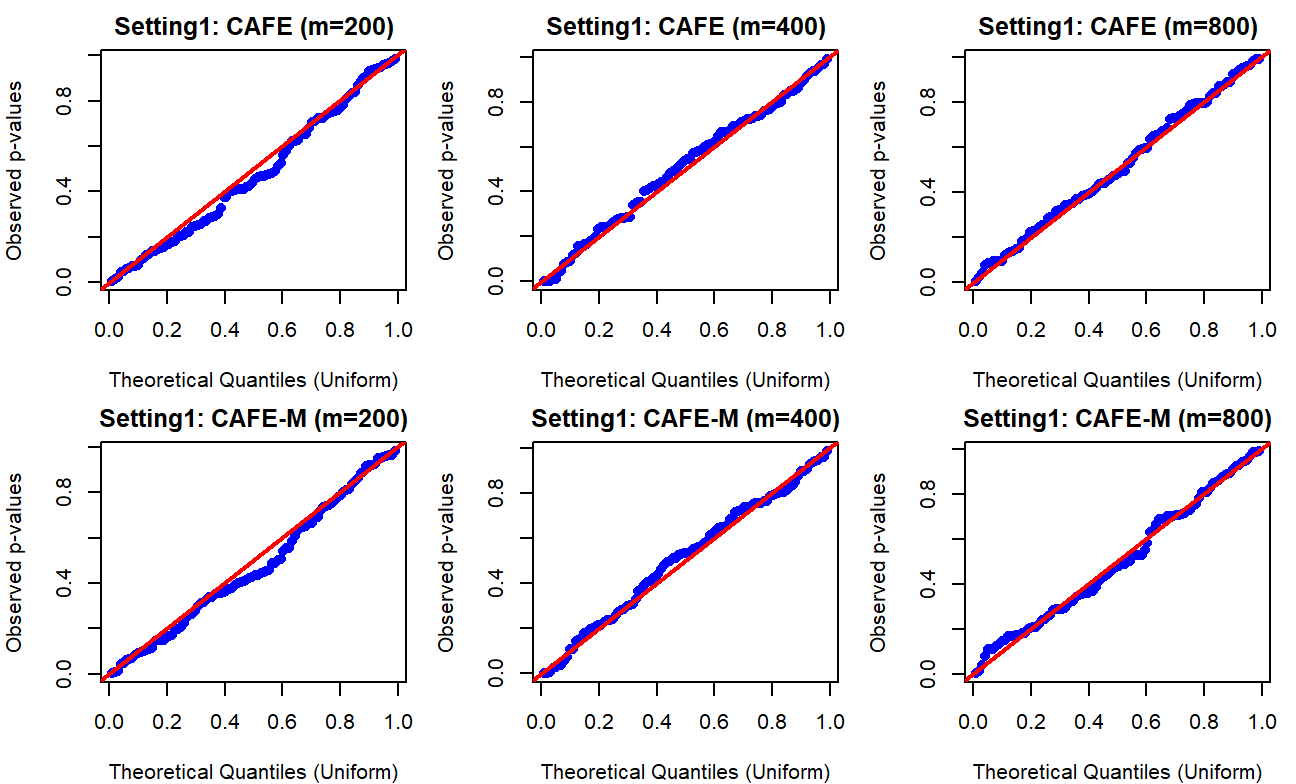}
			\caption{Q--Q plots of the CAFE and CAFE-M p-values under the null hypothesis in Parametric Setting~1. The top row reports CAFE and the bottom row reports CAFE-M. The \(x\)-axis and \(y\)-axis correspond to the theoretical Uniform\((0,1)\) quantiles and the observed sample quantiles, respectively.}
			\label{fig:setting1_TM}
		\end{figure}

		
		Figure~\ref{fig:setting1_TM} reports Q--Q plots of p-values under the null hypothesis in setting 1. The empirical quantiles are close to the theoretical Uniform\((0,1)\) quantiles, indicating satisfactory type I error control for both CAFE and CAFE-M. The agreement improves as the sample size increases. Similar Q--Q plots for Parametric Settings~2 and~3 are reported in the supplementary material and show comparable calibration patterns.
		\begin{figure}[h]
			\centering
			\begin{subfigure}[b]{0.32\textwidth}
				\includegraphics[width=\textwidth]{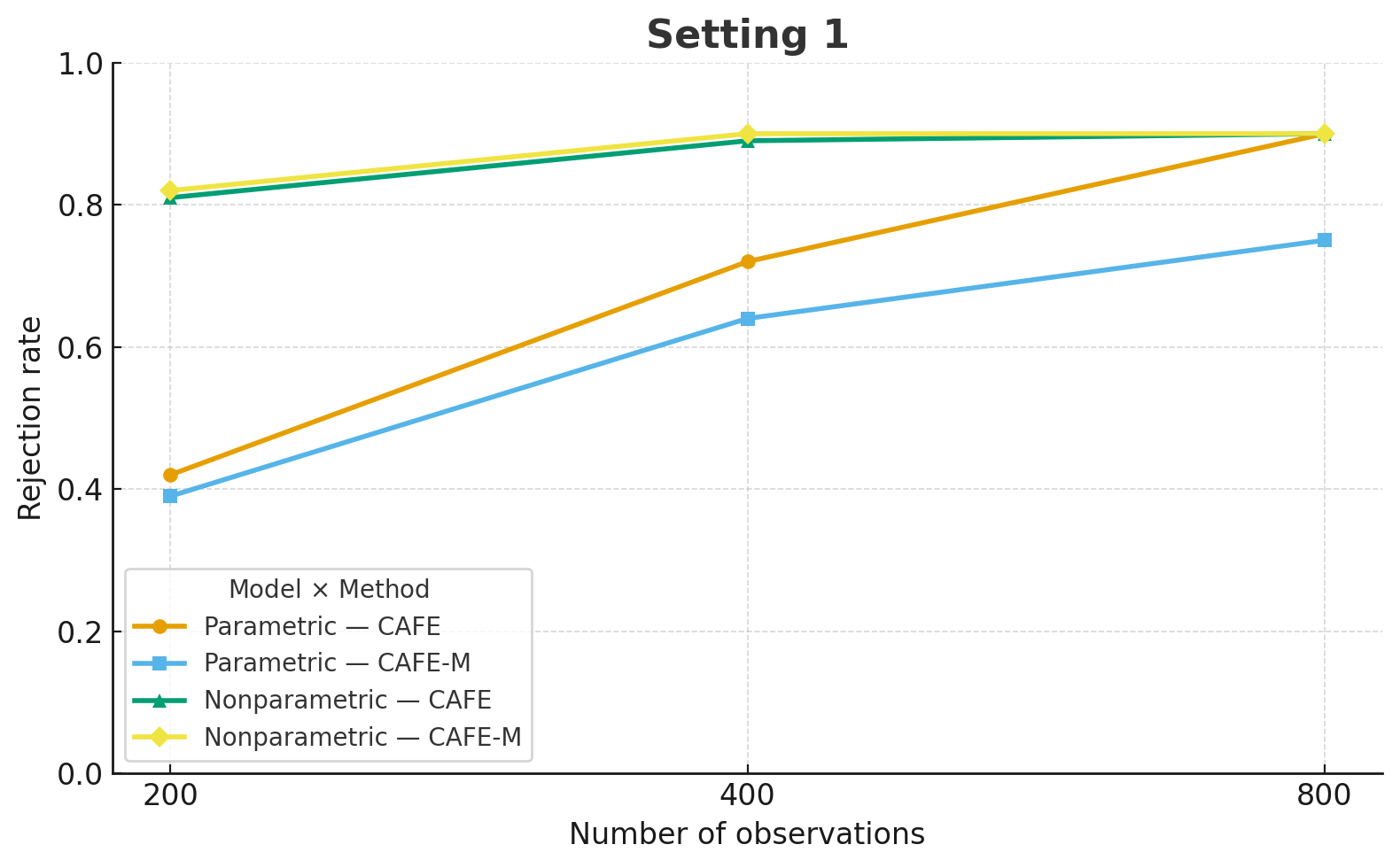}
				\caption{Parametric Setting~1}
			\end{subfigure}
			\begin{subfigure}[b]{0.32\textwidth}
				\includegraphics[width=\textwidth]{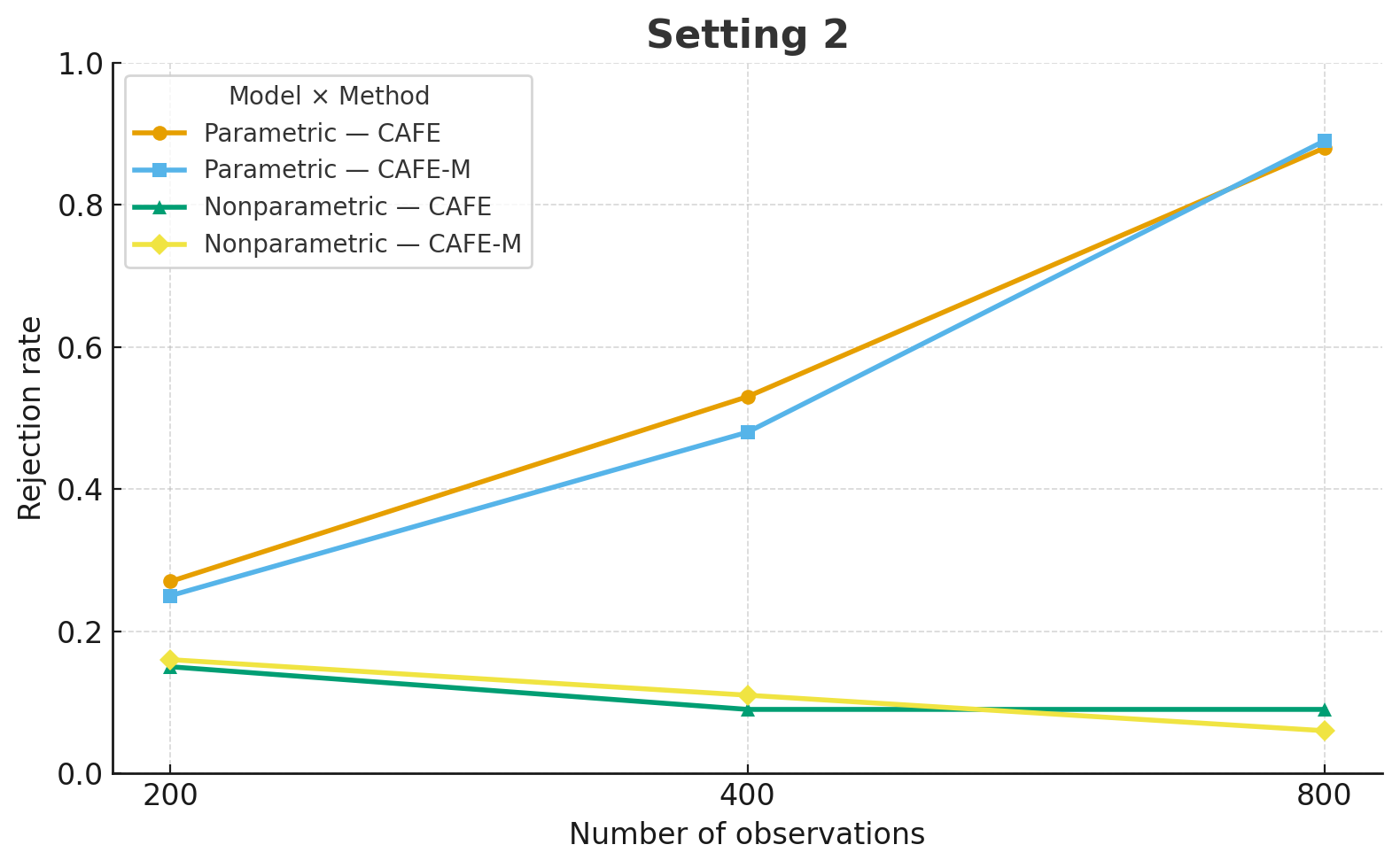}
				\caption{Parametric Setting~2}
			\end{subfigure}
			\begin{subfigure}[b]{0.32\textwidth}
				\includegraphics[width=\textwidth]{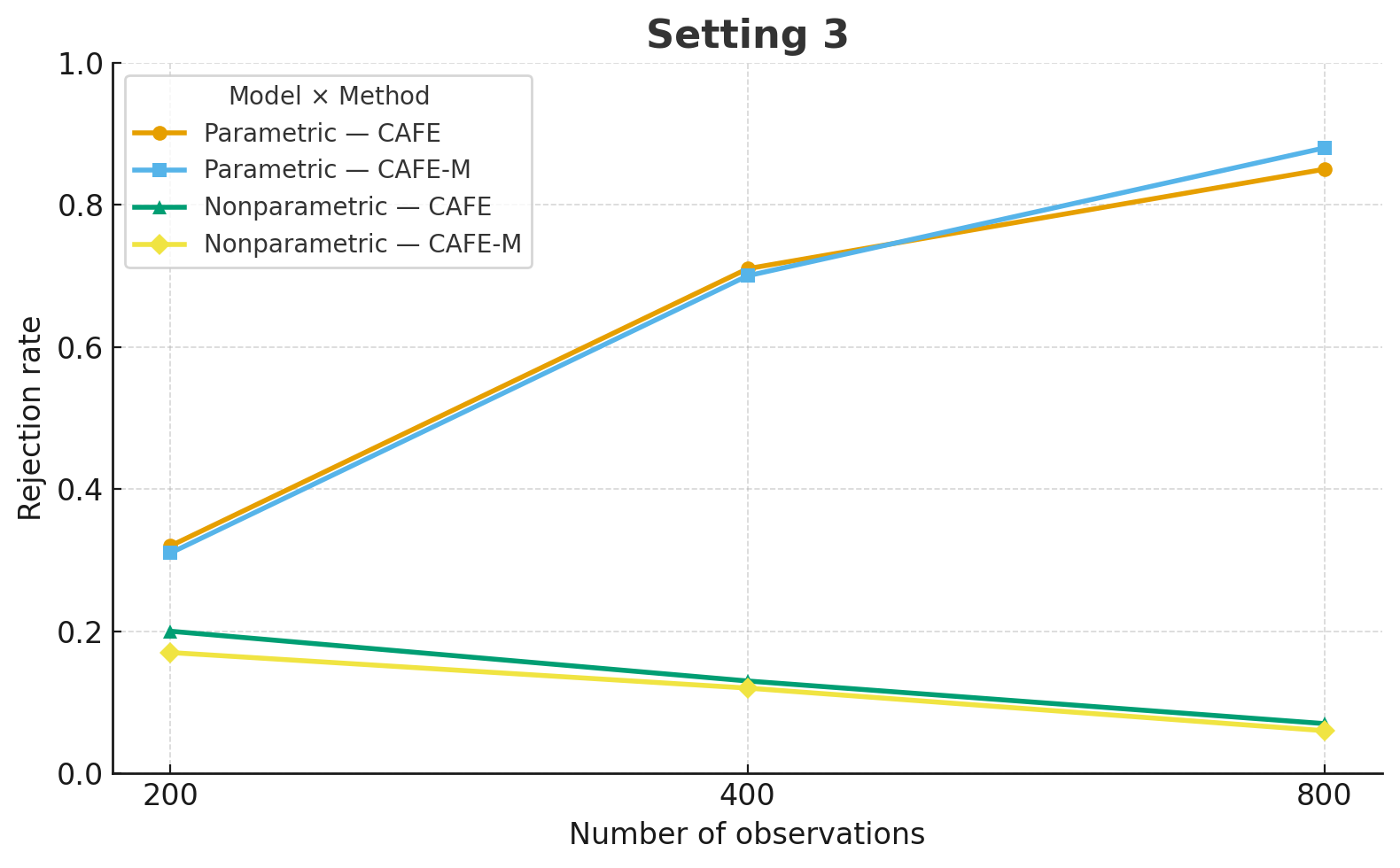}
				\caption{Parametric Setting~3}
			\end{subfigure}
			\caption{Rejection rates of CAFE and CAFE-M under misspecified parametric models and the XGBoost-based R-learner across the three parametric settings.}
			\label{fig:mis}
		\end{figure}
		Figure~\ref{fig:mis} reports empirical rejection rates under the alternative for Parametric Settings~1--3. When the observational CATE is fitted by a misspecified parametric model, both CAFE and CAFE-M show monotone gains in power with the increased observational sample size. The behavior of CAFE and CAFE-M with the XGBoost R-learner is closely tied to the data-generating process. When the true model is linear, the tests tend to reject the null hypothesis; this pattern can be attributed to the slower convergence rate and higher variance of the nonparametric estimator in a simple parametric scenario. By contrast, in the nonlinear designs, the XGBoost R-learner is better able to capture complex relationships, and the resulting test statistics remain close to the nominal level, indicating that the learner is well-suited for nonlinear CATEs.
		\subsection{Enhanced testing with observed confounders or proxies in randomized trial}
		Unobserved confounding is a central concern in observational causal inference, and it cannot be assessed from observational data alone. In practice, randomized trials often collect richer baseline covariates than observational studies. Some of these RCT-only variables may correspond to confounders omitted from the OS, or to informative proxies for them. Incorporating such variables into the partitioning step can improve the ability of CAFE and CAFE-M to detect discrepancies between OS-based CATE estimates and randomized evidence.
		
		We revisit Parametric Setting~1, where \(X_5\) affects both treatment assignment and outcomes but is omitted from the observational CATE and propensity score estimators. We compare three RCT partitioning strategies: partitions based on the estimated propensity score, partitions based on the observational CATE estimate, and partitions based on quantiles of \(X_5\). The latter represents the case where the RCT contains an important covariate that is unavailable or omitted in the observational analysis. We also compare CAFE and CAFE-M with the semiparametric efficiency score (SES) test of \cite{yang2023elastic}, which is designed to detect violations of unconfoundedness in observational studies.
			\begin{figure}[h]
				\centering
				\includegraphics[width=0.75\textwidth]{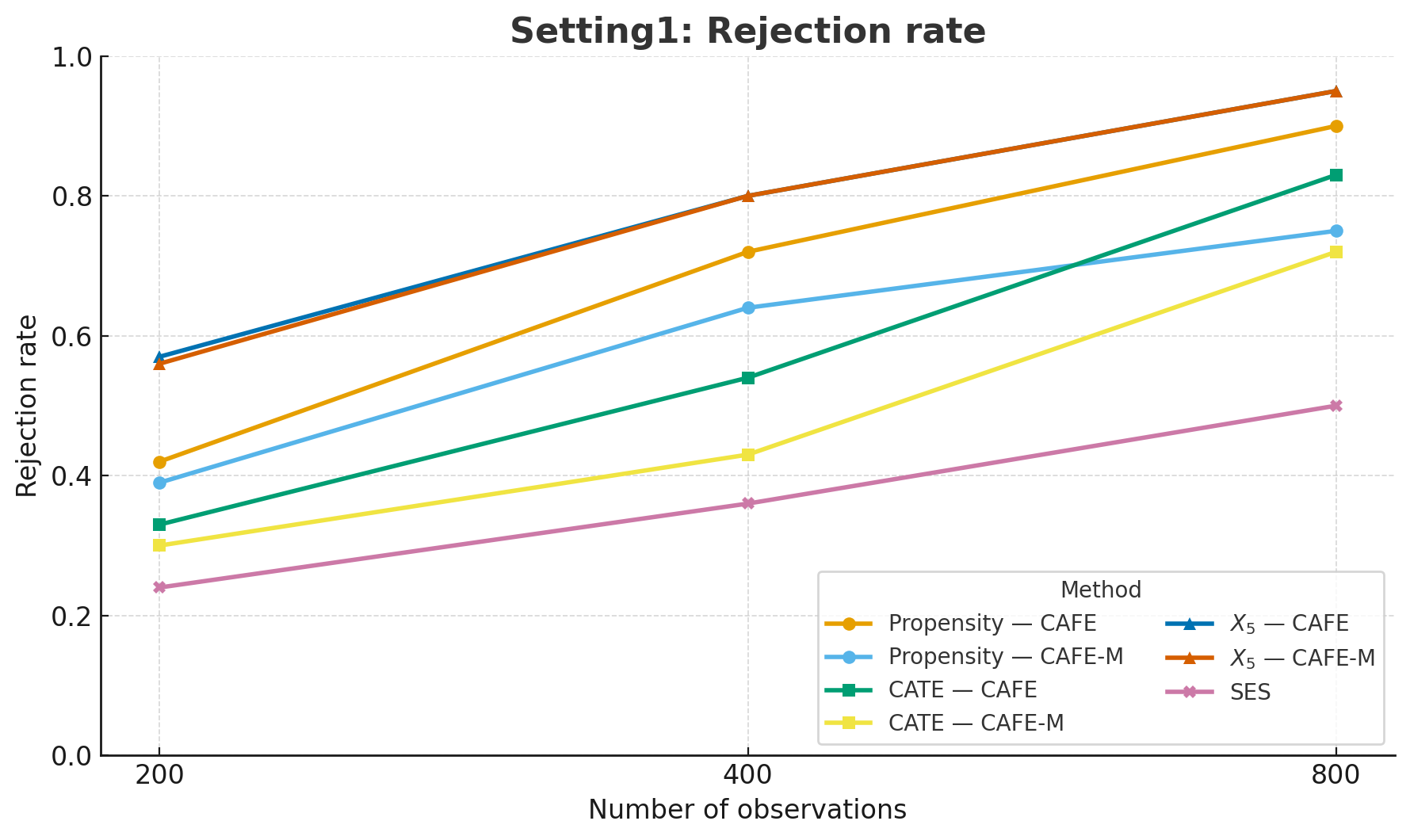}
				\caption{Parametric Setting~1: rejection rates of SES, CAFE and CAFE-M under three partitioning strategies.}
				\label{fig:setting1_x5}
			\end{figure} 
			Figure~\ref{fig:setting1_x5} shows that partitions aligned with the omitted confounder yield the highest rejection rates. In particular, stratifying the RCT by \(X_5\) substantially improves the power of both CAFE and CAFE-M. This is because \(X_5\)-based partitions create groups in which the omitted-variable bias is relatively homogeneous within groups but varies across groups, thereby sharpening the contrast between RCT group-level treatment effects and OS-based predictions. Across sample sizes, CAFE and CAFE-M also achieve higher rejection rates than the SES test. Additional simulations in which the RCT contains only a proxy of an unobserved confounder are reported in the supplementary material.

			\subsection{Is there evidence of unobserved confounding?}
			We now examine the diagnostic procedure from Section~\ref{new sec}. We construct two representative scenarios: one with the source of discrepancy being unobserved confounding, and the other being model misspecification.
			
			For the confounding case, we revisit Parametric Setting~1. The working model is correctly specified for the CATD \(\psi(\mathbf{x})\), but the observational estimator \(\hat{\tau}_O(\mathbf{x})\) omits \(X_5\), which acts as an unobserved confounder. We first test \(\hat{\tau}_O\) against the RCT sample and then assess its fit to the CATD using an OS test set, based on a 90/10 train-test split. For the misspecification case, we revisit Parametric Setting~3. All treatment-effect modifiers are observed, but the working model omits the quadratic term \(x_1^2\). The OS sample is split into 80\% training data and 20\% test data, and the same two-stage testing procedure is applied.
			\begin{figure}[h]
				\centering
				\begin{subfigure}{0.48\textwidth}
					\centering
					\includegraphics[width=\linewidth]{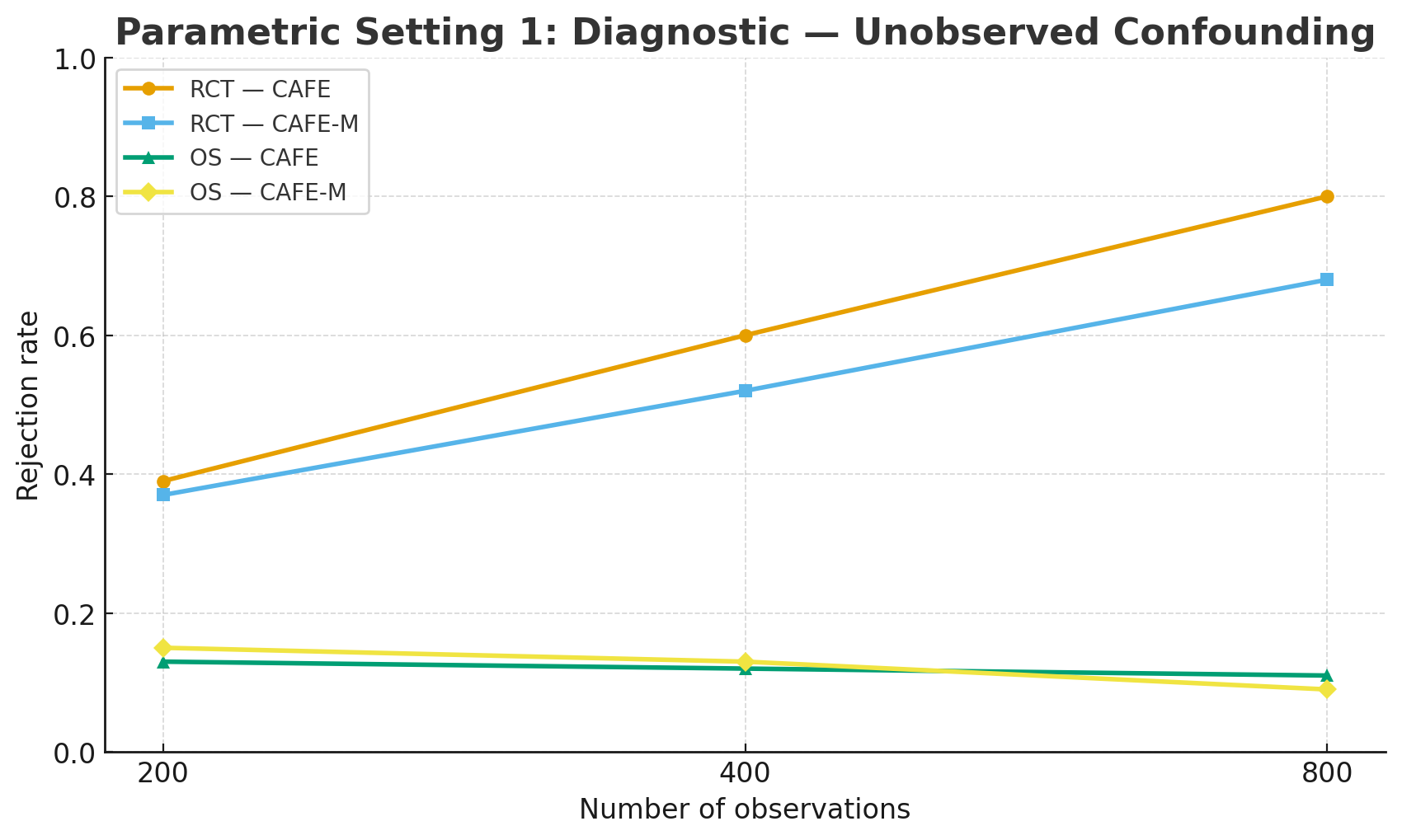}
				\end{subfigure}
				\hfill
				\begin{subfigure}{0.48\textwidth}
					\centering
					\includegraphics[width=\linewidth]{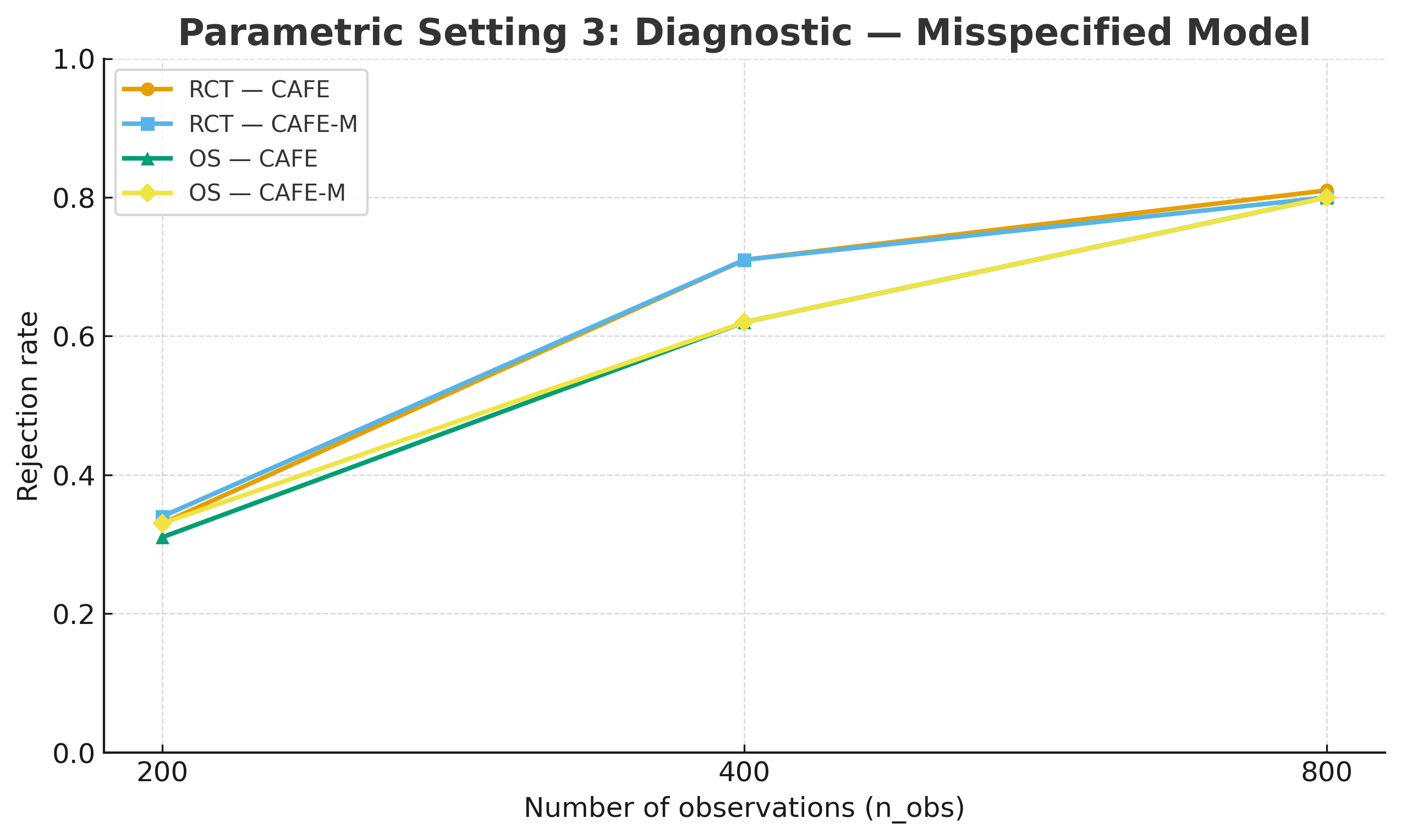}
				\end{subfigure}
				\caption{Rejection rates for CAFE and CAFE-M based on RCT and OS in Parametric Settings~1 and~3.}
				\label{diag}
			\end{figure}
			
			Figure~\ref{diag} summarizes the rejection rates from the two diagnostic tests. In the confounding case, the RCT-based tests reject more frequently as \(m\) increases, whereas the OS-based tests remain close to the nominal level. This pattern suggests that the RCT discrepancy is driven mainly by unobserved confounding rather than poor estimation of the CATD. In the misspecification case, both the RCT-based and OS-based tests reject more frequently as \(m\) grows, indicating that the lack of fit is primarily due to inadequate modeling of the observational contrast. These results illustrate how the two-stage procedure helps distinguish confounding-driven discrepancies from model misspecification.
			\subsection{High-dimensional settings}
			We now turn to high-dimensional designs with $p=1000$ independent covariates, an observational dataset of size $m=500$, and a randomized trial dataset of size $n=120$. Outcomes are generated from,
			\[
			Y(0) = \mu(\mathbf{X}) + \epsilon, \qquad
			Y(1) = \mu(\mathbf{X}) + \tau(\mathbf{X}) + \epsilon,
			\]
			where $\epsilon \sim \mathcal{N}(0,1)$. We consider two scenarios below.
			
			\textit{High-dimensional Setting 1:}  
			\[
			\mu(\mathbf{x}) = \sin(2\pi x_1), \quad
			\tau(\mathbf{x}) = \sin(\pi x_1 x_2) + 2(x_3 - 0.5)^2 + x_4 - 0.5x_5.
			\]
			where the covariates $\mathbf{x} = (x_1, x_2, \ldots, x_{1000})^T$ are independently sampled from $U(-3, 3)$. The observational propensity score is
			\begin{equation} \label{high_e}
				e_0(\mathbf{x})  = \frac{1}{1+\exp(0.5x_1-0.3x_2+0.2x_3+0.1x_4-0.1x_5)}.
			\end{equation}
			The RCT sample is partitioned based on the quantiles of observational CATE estimates $\hat{\tau}_O(\mathbf{x})$. This setting represents a sparse nonlinear treatment effect.
			\\
			\textit{High-dimensional Setting 2:}  The covariates follow \(N(0,I_{1000})\), and
			\[
			\mu(\mathbf{x}) = 1.5x_1 - x_2 + 0.5x_3 + 0.3x_4 - 0.8x_5,
			\]
			\[
			\tau(\mathbf{x}) = x_1 + 0.8x_2 - 0.6x_3 + 0.4x_4 - 1.2x_5.
			\]
			The observational propensity score follows the same logistic specification as (\ref{high_e}), and the RCT is partitioned by quantiles of the estimated propensity score.

			We evaluate CAFE and CAFE-M using several representative CATE estimators, including Lasso-based and XGBoost-based R-learners and T-learners \citep{tibshirani1996regression, chen2016xgboost}, as well as causal forest \citep{wager2018estimation}. For all methods requiring propensity score estimation, Lasso logistic regression is used.
			\begin{figure}[htbp]
				\centering
				
				\begin{subfigure}{0.48\textwidth}
					\centering
					\includegraphics[width=\linewidth]{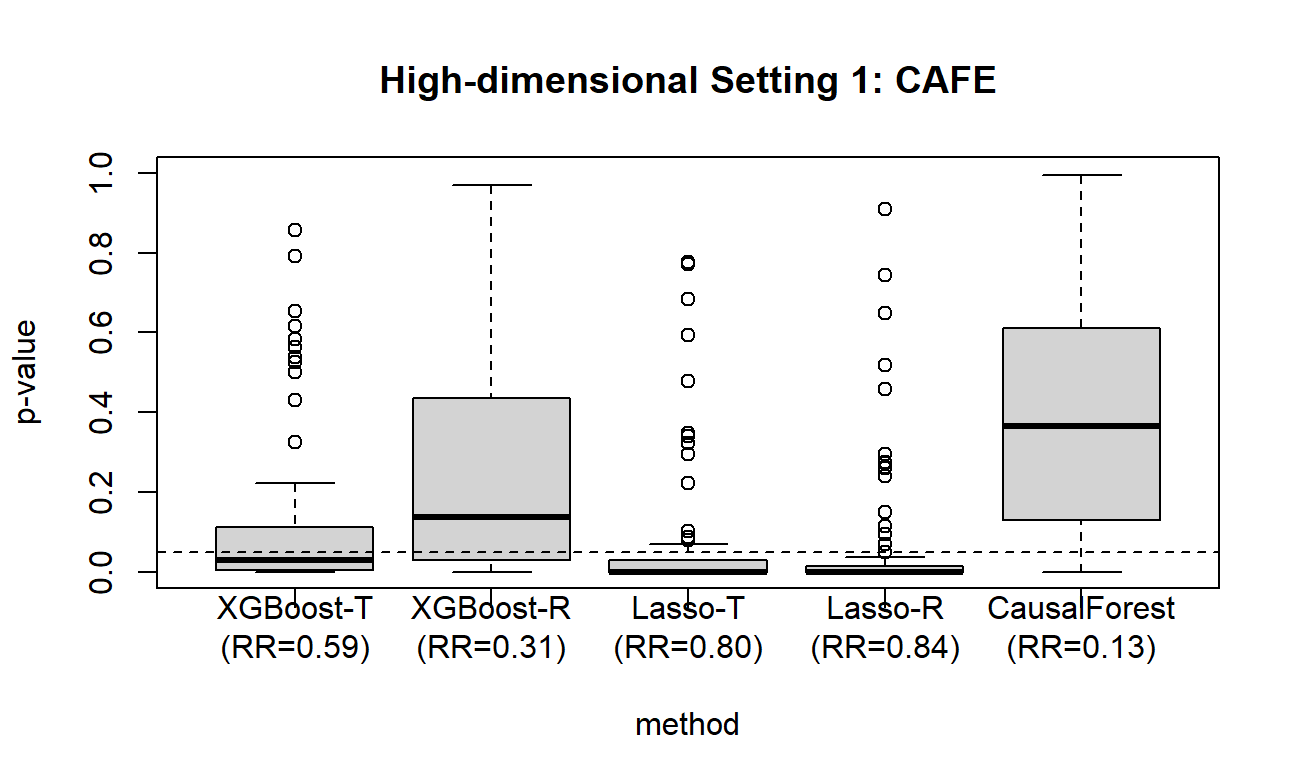}
					\caption{High-dimensional Setting~1: CAFE}
					\label{fig:hd1_cafe}
				\end{subfigure}\hfill
				\begin{subfigure}{0.48\textwidth}
					\centering
					\includegraphics[width=\linewidth]{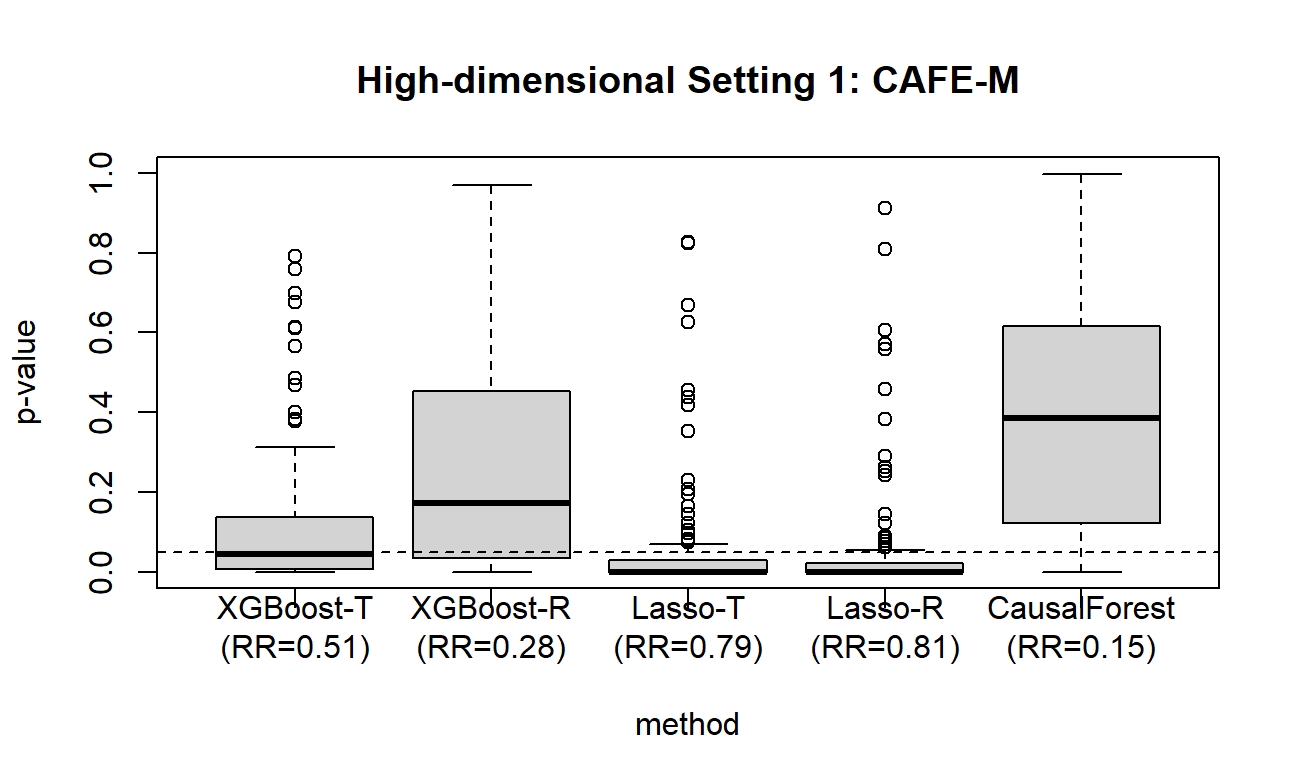}
					\caption{High-dimensional Setting~1: CAFE-M}
					\label{fig:hd1_cafem}
				\end{subfigure}
				\medskip
				
				\begin{subfigure}{0.48\textwidth}
					\centering
					\includegraphics[width=\linewidth]{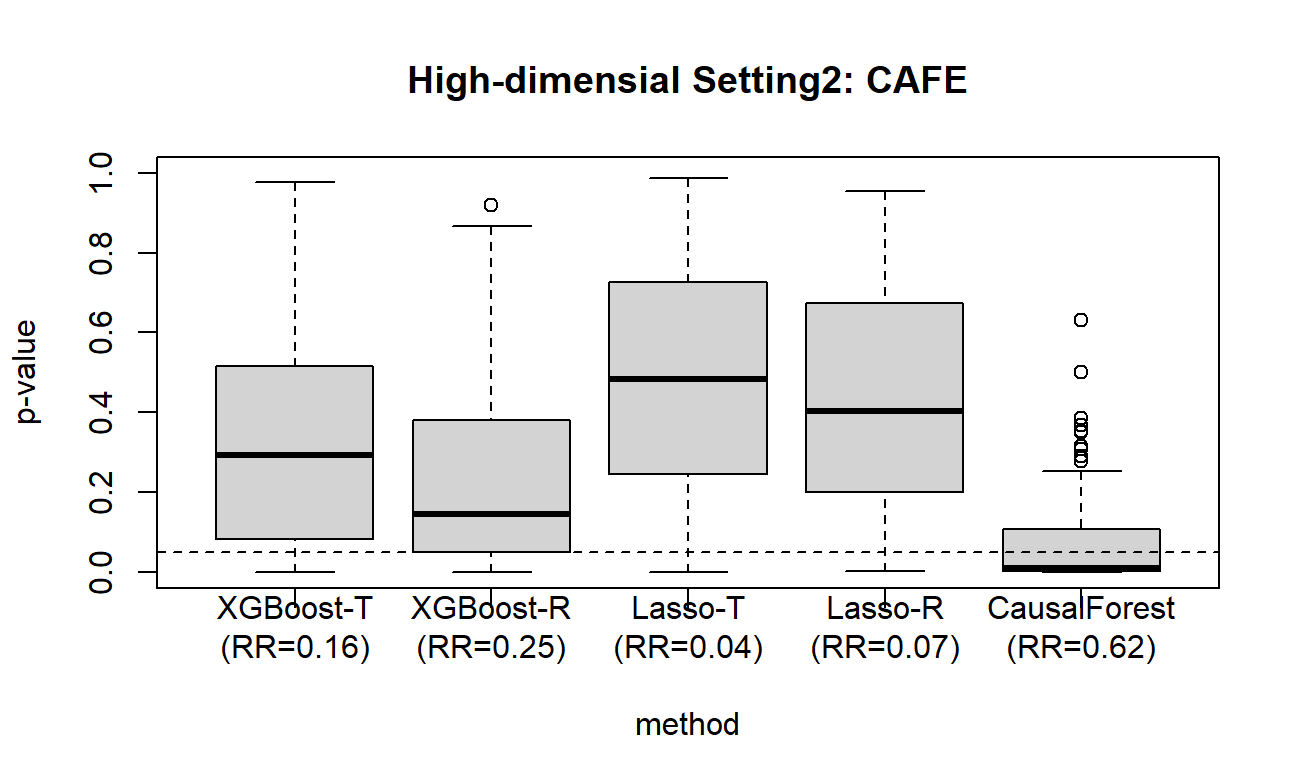}
					\caption{High-dimensional Setting~2: CAFE}
					\label{fig:hd2_cafe}
				\end{subfigure}
				\hfill
				\begin{subfigure}{0.48\textwidth}
					\centering
					\includegraphics[width=\linewidth]{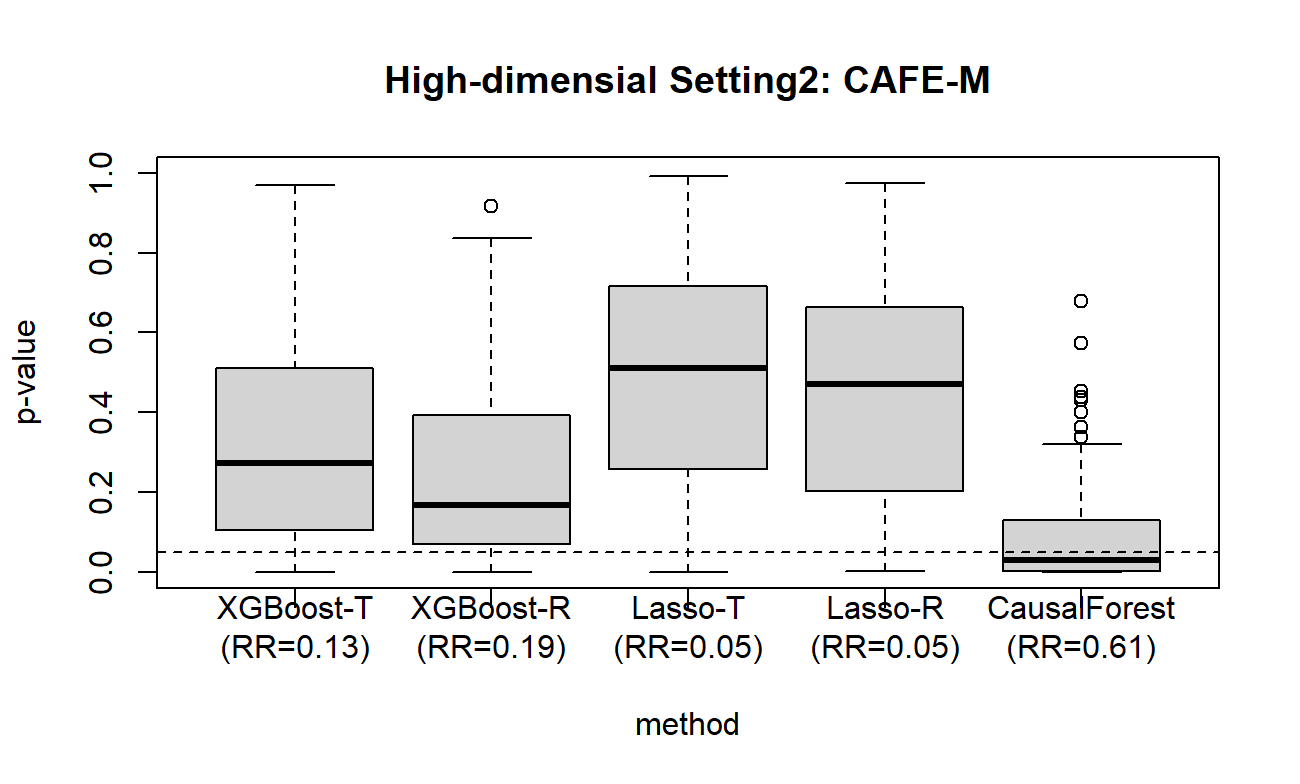}
					\caption{High-dimensional Setting~2: CAFE-M}
					\label{fig:hd2_cafem}
				\end{subfigure}
				
				\caption{High-dimensional settings: distributions of p-values across learners. Horizontal dashed lines indicate the 0.05
					significance level. RR is
					short for {rejection rate}.}
				\label{fig:hd}
			\end{figure}
			
		Figure~\ref{fig:hd} summarizes the p-values by boxplots and reports the empirical rejection rates for CAFE and CAFE-M across different CATE learners. In High-dimensional Setting 1, where the true outcome functions are nonlinear and smooth, causal forest performs best among the methods considered. The XGBoost-based R-learner is moderately better than XGBoost-based T learner, whereas the Lasso-based learners are deemed unfit with high probabilities, as expected under nonlinear model misspecification. In High-dimensional Setting 2, where the CATE is linear and sparse, the Lasso-based learners show rejection rates very close to the nominal level, whereas the XGBoost-based learners are rejected around 20\% because $m=500$ is not large enough for them to perform stably well. By contrast, causal forest performs very poorly, suggesting that causal forest is much less stable in this sparse linear high-dimensional setting, likely due to slower adaptation to the underlying sparse structure.
			\section{Data analysis} \label{sec7}
		We analyze data from the Tennessee Student/Teacher Achievement Ratio (STAR) experiment \citep{word1990state, krueger1999experimental}, a large-scale randomized trial initiated in 1985 to study the effect of class size on student achievement. The dataset has been widely used in heterogeneous treatment effect estimation and in studies combining randomized and observational data \citep{kallus2018removing, wu2022integrative}. We use STAR to construct three controlled settings that examine, respectively, baseline performance under unconfoundedness, sensitivity to model misspecification, and sensitivity to hidden confounding.
			
			Our analysis focuses on students who entered the study in first grade and were assigned to either small classes (13–17 students) or regular-sized classes (22–25 students). Of the 4,509 students initially randomized, we exclude those with missing outcomes, yielding a final sample of 4,218 students: 1,805 in the treatment group (small classes) and 2,413 in the control group (regular classes). The outcome $Y$ is defined as the composite standardized score in listening, reading, and mathematics at the end of first grade. Covariates $\mathbf{X}$ include gender, race, birth date, free-lunch eligibility, teacher identifiers, and rural/inner-city indicators.
			
			We first construct an unconfounded setting by randomly splitting the STAR sample into RCT and observational samples in a 10:90 ratio. Since treatment assignment in STAR was randomized, this split preserves unconfoundedness. We fit four observational CATE learners: linear regression, ridge regression, causal forest, and an XGBoost-based R-learner. For each fitted learner, the RCT sample is partitioned by quantiles of the CATE estimates, and both CAFE and CAFE-M are implemented with \(K=3,4,5\).
			
			To assess sensitivity to model misspecification, we omit race from the fitted CATE model and modify the outcome by adding a race-related treatment-effect component, \[
			Y_{\mathrm{new}} = Y + 0.5(2T-1)\,\tilde{X}_R,
			\]
		where \(\tilde{X}_R\) is a standardized race contrast score. Because this perturbation enters through the treatment contrast, omitting race creates a controlled omitted-effect-modifier misspecification. To assess sensitivity to hidden confounding, we redraw treatment assignments in the observational sample from a logistic propensity score depending on individual baseline covariates and school-level baseline composition measures. The school-level quantities affect treatment assignment but are excluded from the covariates used to fit the CATE learners, and hence act as latent baseline confounders from the perspective of the observational analysis.
			
			\begin{table}[htbp]
				\centering
				\caption{Average $p$-values for CAFE and CAFE-M under unconfoundedness }
				\label{tab:cafe_results}
				\begin{tabular}{llccc}
					\toprule
					Model & Test & $K=3$ & $K=4$ & $K=5$ \\
					\midrule
					\multirow{2}{*}{Linear Regression}
					& CAFE   & $<0.0001$ & $<0.0001$ & $<0.0001$ \\
					& CAFE-M & $<0.0001$ & $<0.0001$ & $<0.0001$ \\
					\midrule
					\multirow{2}{*}{Ridge Regression}
					& CAFE   & 0.3122 &0.3005 & 0.3096 
					\\
					& CAFE-M & 0.3234 & 0.2940 & 0.3232 \\
					\midrule
					\multirow{2}{*}{Causal Forest}
					& CAFE & 0.4551 & 0.4172 & 0.4218 \\
					& CAFE-M & 0.4485 & 0.4127 & 0.4191 \\
					\midrule
					\multirow{2}{*}{XGBoost R-learner}
					& CAFE & 0.0262 & 0.0209 & 0.0235 \\
					& CAFE-M & 0.0304 & 0.0244 & 0.0300 \\
					\bottomrule
				\end{tabular}
			\end{table}
			\begin{table}[htbp]
				\centering
				\caption{Average $p$-values for CAFE and CAFE-M under  misspecification (omitting race)}
				\label{tab:cafe_missing}
				\begin{tabular}{llccc}
					\toprule
					Model & Test & $K=3$ & $K=4$ & $K=5$ \\
					\midrule
					\multirow{2}{*}{Linear Regression} & CAFE   & $<0.0001$ & $<0.0001$ & $<0.0001$ \\
					& CAFE-M & $<0.0001$ & $<0.0001$ & $<0.0001$ \\
					\midrule
					\multirow{2}{*}{Ridge Regression}  & CAFE    & 0.0001 & $<0.0001$ & $<0.0001$ \\
					& CAFE-M  & 0.0001 & 0.0002 & $<0.0001$ \\
					\midrule
					\multirow{2}{*}{Causal Forest} &  CAFE & 0.0307 & 0.0187 & 0.0189 \\
					& CAFE-M & 0.0442 & 0.0289 & 0.0277 \\
					\midrule
					\multirow{2}{*}{XGBoost R-learner}  & CAFE   & $<0.0001$ & $<0.0001$ & $<0.0001$ \\
					& CAFE-M & $<0.0001$ & $<0.0001$ & $<0.0001$ \\
					\bottomrule
				\end{tabular}
			\end{table}
			\begin{table}[htbp]
				\centering
				\caption{Average $p$-values for CAFE and CAFE-M with confoundedness }
				\label{tab:cafe_confoud}
				\begin{tabular}{llccc}
					\toprule
					Model & Test & $K=3$ & $K=4$ & $K=5$ \\
					\midrule
					\multirow{2}{*}{Linear Regression} & CAFE & 0.0413 & 0.0492 & 0.0405 \\
					& CAFE-M & 0.0549 & 0.0676 & 0.0633 \\
					\midrule
					\multirow{2}{*}{Ridge Regression}  & CAFE    & 0.0046 & 0.0091 & 0.0060 \\
					& CAFE-M & 0.0098 & 0.0159 & 0.0140 \\
					\midrule
					\multirow{2}{*}{Causal Forest} & CAFE  &  0.0074 & 0.0137 & 0.0124 \\
					& CAFE-M & 0.0164 & 0.0296 & 0.0297 \\
					\midrule
					\multirow{2}{*}{XGBoost R-learner} & CAFE & 0.0184 & 0.0292 & 0.0277 \\
					& CAFE-M & 0.0316 & 0.0496 & 0.0487 \\
					\bottomrule
				\end{tabular}
			\end{table}
			
			Tables~\ref{tab:cafe_results}--\ref{tab:cafe_confoud} report the average
			$p$-values over 100 replicates. In the unconfounded benchmark setting
			(Table~\ref{tab:cafe_results}), ridge regression and causal forest yield relatively
			large average $p$-values, whereas linear regression and the XGBoost R-learner are
			rejected by both CAFE and CAFE-M.
			Under model misspecification (Table~\ref{tab:cafe_missing}), all four learners are
			rejected after a race-related treatment-effect component is introduced but race is
			excluded from the fitted CATE model. Under hidden confounding
			(Table~\ref{tab:cafe_confoud}), all methods yield small average $p$-values.
			
			Overall, these results show that CAFE and CAFE-M provide informative group-level
			diagnostics for observationally trained CATE estimates, with sensitivity to both
			omitted effect modifiers and hidden confounding.
			
			\section{Conclusion and discussion} \label{sec8}
			As pointed out by \cite{rolling2014model}, conditional treatment effect assessment may differ substantially from evaluation of the overall model. In this work, when an independent RCT dataset is available, our approach can directly assess the goodness-of-fit of a CATE estimate based on an observational study.
			The proposed tests remain valid without imposing parametric assumptions on the CATE and can be applied to flexible, black-box CATE estimators. Furthermore, when both RCT and OS data are available, our two-stage testing procedure provides a principled way to investigate the source of an observed lack of fit, helping to distinguish model misspecification from hidden confounding.
			
				The framework may also be adapted to settings where the OS and RCT do not record exactly the same covariates. If the RCT contains additional covariates unavailable in the OS, our test is clearly applicable to evaluate the adequacy of the CATE estimate from the OS, although the failure of rejection should not be interpreted as ruling out effect modification by RCT-only variables. In this case, we may also try using the RCT-only variables for partition which may improve the power of the tests if some RCT-only variables are involved. If the OS contains covariates unavailable in the RCT, the fitted CATE function can only be evaluated after these variables are supplied or imputed for RCT units, and the resulting test should be interpreted under the chosen imputation rule. We provide additional implementation details for these covariate-mismatch settings in the supplementary material.
			
			More broadly, similar ideas of CAFE can be applied when OS and RCT data are jointly used for estimation, provided that proper data splitting is used and the RCT observations used for evaluation are kept independent of those used for model construction. For example, one may split the RCT sample into training and testing subsets, use the former together with the OS to construct the estimator, and reserve the latter exclusively for evaluation by CAFE or CAFE-M. This separation is essential for preserving the validity of the goodness-of-fit assessment.

			\bibliographystyle{abbrvnat}
			\bibliography{JRSSB-main-manuscript}

\end{document}